\documentclass[twocolumn,prl,aps,showpacs,floatfix,superscriptaddress]{revtex4}
\usepackage{amsmath}
\usepackage{amssymb}
\usepackage[dvips]{graphicx}
\usepackage{dcolumn}% Align table columns on decimal point
\usepackage{bm}% bold math
\usepackage{ifpdf}
\usepackage{color}
\usepackage{multirow}

\newcommand{\ZnO}{MgZnO/ZnO}
\newcommand{\mbor}{$\mu^*_{\text{B}}$}
\newcommand{\meff}{$m^*$}

\begin{document}

\title{Enhanced quantum oscillatory magnetization and non-equilibrium currents in an interacting two-dimensional electron system in MgZnO/ZnO with repulsive scatterers}
%Enhanced quantum oscillatory magnetization in an oxide heterostructure due to electron-electron interaction and disorder by repulsive scatterers}% Force line breaks with \\

\author{M. Brasse}
\affiliation{Lehrstuhl f\"{u}r Physik funktionaler Schichtsysteme, Technische Universit\"{a}t M\"{u}nchen, 85748 Garching, Germany}

\author{S.~M. Sauther}
\affiliation{Lehrstuhl f\"{u}r Physik funktionaler Schichtsysteme, Technische Universit\"{a}t M\"{u}nchen, 85748 Garching, Germany}

\author{J. Falson}
\affiliation{Department of Applied Physics and Quantum-Phase Electronics Center (QPEC), University of Tokyo, Tokyo 113-8656, Japan}

\author{Y. Kozuka}
\affiliation{Department of Applied Physics and Quantum-Phase Electronics Center (QPEC), University of Tokyo, Tokyo 113-8656, Japan}

\author{A. Tsukazaki}
\affiliation{Institute for Materials Research, Tohoku University, Sendai 980-8577, Japan, and PRESTO, Japan Science and Technology Agency (JST), Tokyo 102-0075, Japan}

\author{Ch. Heyn}
\affiliation{Institut f{\"u}r Angewandte Physik, Universit{\"a}t Hamburg, Jungiusstrasse 11, 20355 Hamburg, Germany}

\author{M.~A. Wilde}
\email[]{mwilde@ph.tum.de}
\affiliation{Lehrstuhl f\"{u}r Physik funktionaler Schichtsysteme, Technische Universit\"{a}t M\"{u}nchen, 85748 Garching, Germany}

\author{M. Kawasaki}
\affiliation{Department of Applied Physics and Quantum-Phase Electronics Center (QPEC), University of Tokyo, Tokyo 113-8656, Japan}

\author{D. Grundler}
\affiliation{Lehrstuhl f\"{u}r Physik funktionaler Schichtsysteme, Technische Universit\"{a}t M\"{u}nchen, 85748 Garching, Germany}

\date{\today}% It is always \today, today,

\begin{abstract}
Torque magnetometry at low temperature and in high magnetic fields $B$ is performed on a {\ZnO} heterostructure incorporating a high-mobility two-dimensional electron system. We find a sawtooth-like quantum oscillatory magnetization $M(B)$, i.e., the de Haas-van Alphen (dHvA) effect. At the same time, unexpected spike-like overshoots in $M$ and non-equilibrium currents are observed which allow us to identify the microscopic nature and density of the residual disorder. The acceptor-like scatterers give rise to a magnetic thaw down effect which enhances the dHvA amplitude beyond the electron-electron interaction effects being present in the {\ZnO} heterostructure.
\end{abstract}
\pacs{Valid PACS appear here}

\maketitle

\begin{figure}[htbp]
\center
\includegraphics[width=7cm]{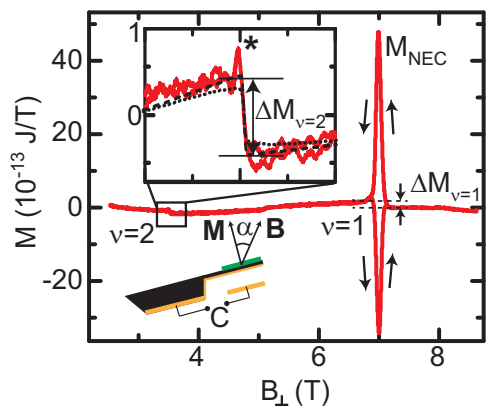}
\caption{(Color online) Oscillatory part of the magnetization of a MgZnO/ZnO heterostructure at $\alpha=52.0^{\circ}$ and $T=280$~mK. At filling factor $\nu=1$, the data exhibits the dHvA effect as well as large NECs, which change sign upon changing the sweep direction (arrows). In the inset, experimental data (light line) as well as model calculations for the ideal (dashed line) and real (dotted line) 2DES are shown for $T=280$~mK ($3.0~{\rm T}<B<4.0~{\rm T}$).}
\label{fig1}
\end{figure}

\begin{figure}[htbp]
\center
\includegraphics[width=7cm]{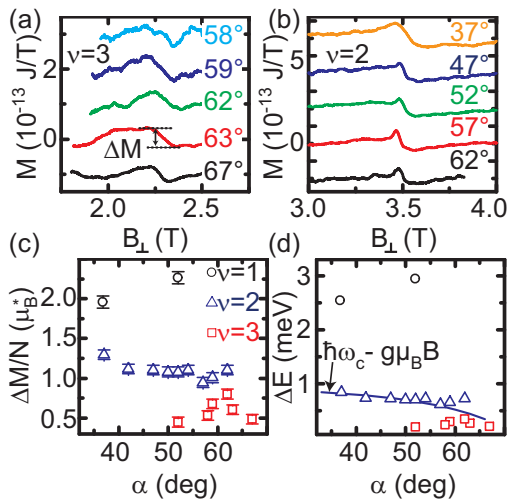}
\caption{(Color online) Angular dependencies of $M(B)$ near (a) $\nu=3$ and (b) $\nu=2$. Curves are shifted in vertical direction for clarity. (c) DHvA amplitudes per electron $\Delta M_{\text{e},\nu}=\Delta M_{\nu}/N$ as a function of $\alpha$. Values for $\nu=1$ were extracted from curves shown in Fig.~\ref{fig1} and Fig.~\ref{fig3}~(a), respectively. (d) Thermodynamic energy gaps $\Delta E_{\nu}$ as a function of $\alpha$. The line denotes the energy gap expected for $\nu=2$ in a disorder-free 2DES at $T=0$.}
\label{fig2}
\end{figure}

Oxide heterostructures have generated tremendous interest in recent years \cite{Othomo2004,Hwang2012}. Two-dimensional electron systems (2DESs) formed therein exhibit remarkable properties such as superconductivity \cite{Reyren2007} and magnetism \cite{Bert2011} or the fractional quantum Hall effect (QHE) \cite{Tsukazaki2007,Tsukazaki2010}. MgZnO/ZnO-based heterostructures are outstanding in that 2DESs of small carrier density $n_s$ exhibit extremely high mobilities $\mu$ at low temperature $T$ \cite{Tsukazaki2007,Tsukazaki2010}. At the same time, the electron-electron interaction parameter $r_{\text{s}}\propto n_s^{-0.5}$ \cite{Fulde2012} is large allowing for electron correlation effects at oxide interfaces in an applied magnetic field $B$ \cite{Tsukazaki2008,Tsukazaki2010,Kozuka2012}. Still, the quantum oscillatory magnetization $M(B)$, i.e., the de Haas-van Alphen (dHvA) effect reflecting the ground state properties of such 2DESs has not yet been explored. Since the discovery of the dHvA effect in Bi more than eight decades ago it has been argued that disorder {\em reduces} peak-to-peak amplitudes $\Delta M$ via broadening of the quantized Landau levels $E_j$ ($j=0,1,2,...$) \cite{Kittel,Shoenberg1984,Eisenstein1985}. In contrast, electron-electron interaction effects are known to enhance $\Delta M$ \cite{MacDonald1986}. The two counteracting effects are however not easy to distinguish in a balancing situation. Sometimes the dHvA effect has been obscured in the QHE regime even completely by extremely large non-equilibrium currents (NECs) \cite{usher2009}. The NECs are induced near integer filling factors $\nu=hn_{\rm s}/(eB_{\perp})$ at low $T$ when the longitudinal resistivity $\rho_{xx}$ takes a vanishingly small value and are believed to be limited only by breakdown of the QHE \cite{usher2009}. Independent transport experiments on GaAs-based heterostructures have evidenced that, strikingly, minima in $\rho_{xx}$ and plateaus in the Hall resistivity $\rho_{xy}$ could be displaced away from integer $\nu$ towards smaller $\nu$ due to repulsive scatterers \cite{Furneaux1986,Haug1987,Bonifacie2006,Raymond2009,Bisotto2012}. This phenomenon has not yet been resolved in $M(B)$, and a clear experimental manifestation of the underlying asymmetric density of states (DOS) in a ground state property is still lacking.\\ \indent In this Letter, we report torque magnetometry on the equilibrium dHvA effect and NECs of a high-mobility 2DES residing at a \ZnO~heterointerface. We observe dHvA amplitudes $\Delta M$ at filling factors $\nu=1$ and $2$ that are significantly enhanced over the expected values in the single-particle picture. Addressing a regime $0.28~{\rm K}<T<1.6~$ K we observe $T$-dependent shifts of both the dHvA effect and the NECs. At the same time, spike-like overshoots of the equilibrium magnetization are found near filling factors $\nu=1$ and $2$. These unforeseen characteristics of $M$ are consistently attributed to a disorder-induced asymmetric DOS due to repulsive scatterers and the magnetic thaw down of electrons in magnetoacceptor (MA) states \cite{Haug1987,Raymond2009,Bisotto2012}. Contrary to the orthodox thinking, the disorder in \ZnO~is found to provoke thereby an {\em increase} of the equilibrium magnetization near integer $\nu$ instead of the anticipated reduction. In particular, $M(B)$ allows us to quantify the density of MA states. The microscopic understanding of disorder is important to further optimize oxide heterostructures for basic research and technological applications.\\ \indent
The magnetization experiments were performed on a MgZnO/ZnO heterostructure. Details of sample preparation can be found in Refs. \cite{supmaterial,Falson2011}. We investigated two samples from the same heterostructure. The results were consistent. We focus here on the results of the sample which exhibits larger signals. Results obtained on the reference sample can be found in Ref.~\cite{supmaterial}. The electron density $n_{\text{s}}=1.7\times10^{11}$~cm$^{-2}$ and mobility $\mu=380,000~{\rm cm}^2/$Vs were determined from Shubnikov-de Haas and Hall effect measurements on a reference 2DES at $0.5$~K \cite{supmaterial}. $r_{\rm s}$ given by $m_{\text{b}}e^2/(4\pi\epsilon\hbar^2\sqrt{\pi n_{\text{s}}})$, amounted to 9.0, where $e$ denotes the elementary charge, $m_{\text{b}}=0.29m_{\text{e}}$ the band mass ($m_{\text{e}}$ is the free electron mass) and $\epsilon=8.5$ the dielectric constant of ZnO \cite{Maryenko2012}. The area $A$ of the studied 2DES was $0.9\times1.8$~mm$^2$. We employed micromechanical cantilever magnetometry to measure the anisotropic magnetization $\textbf{\text{M}}$ of the 2DES. The sensor was prepared from an undoped AlGaAs/GaAs heterostructure grown by molecular-beam epitaxy \cite{Schwarz2000}. The sample was glued to the flexible end of the cantilever beam as depicted in the inset of Fig.~\ref{fig1}. The deflection was monitored by measuring the capacitance between the beam and a fixed ground plane \cite{Wilde2008}. After calibration following Ref. [\onlinecite{Schwarz2000}] the capacitance change provided the torque $\boldsymbol{\tau}=\mathbf{M}\times\mathbf{B}$. From $\tau/(B\sin\alpha)$ we extracted the magnetization $M$ assuming $\mathbf{M}$ to be perpendicular to the plane given by the 2DES.
We performed the measurements in a $^3$He cryostat at temperatures down to 0.28 K. Two sets of experiments were performed. In a vector magnet system we applied a magnetic field $\mathbf{B}$ of up to 9~T under different angles $36.8^{\circ}<\alpha<67^{\circ}$ in the same cool-down cycle. In an axial magnet, we performed further magnetometry measurements at two fixed angles $\alpha=36.8^{\circ}$ and $\alpha=52^{\circ}$ up to $B=14$~T. To analyze the data we calculated $M=-\partial F/\partial B\vert_{T,n_{\text{s}}}$ from the free energy $F$ considering the single-particle DOS according to \cite{Wilde2008}
\begin{equation}
D(E)=\frac{eB_{\perp}}{h}\sum_{j=0}^{\infty}\frac{1}{\sqrt{2\pi}\Gamma}\exp\left[\frac{(E-E_{\text{j,s}})^2}{2\Gamma^2}\right].
\label{Eq1}
\end{equation}
Here, we use that the energy spectrum of the 2DES is composed of discrete levels $E_{\text{j}}=(j+1/2)\hbar\omega_{\text{c}}$ due to the perpendicular magnetic field component $B_{\perp}=B\cos \alpha$. The cyclotron frequency is given by $\omega_{\text{c}}=eB_{\perp}/m^*$, where $m^*$ denotes the effective electron mass. Each Landau level $E_j$ undergoes spin splitting such that the energy spectrum takes the form $E_{\text{j,s}}=E_{\text{j}}+sg\mu_{\text{B}}B$ with $s=\pm1/2$ and the appropriate Land{\'e} factor $g$. For $T=0$, $M(B)$ of an ideal 2DES exhibits steps of height $\Delta M/N=\Delta M/(n_{\rm s}A)=\Delta E/B$ when the Fermi energy crosses an energy gap $\Delta E$ between two adjacent energy levels at integer filling factors $\nu$ \cite{MacDonald1986,Wiegers1997}. $\Delta M_{\nu}$ is thus a measure of the so-called thermodynamic energy gap $\Delta E$ at $\nu$. In contrast to excitation spectroscopy (e.g. by transport measurements), one does not need to consider microscopic parameters of scattering processes or selection rules for calculating $M$. We took both Gaussian level broadening by $\Gamma=\frac{e\hbar}{m^*}\sqrt{2/\pi\mu}\sqrt{B_{\perp}}$ \cite{Ando1974} and a finite temperature $T$ into account when calculating $F$ and $M$ \cite{Wilde2008}. Measuring $M(B)$ the ground state energy spectrum is probed in a contactless configuration and thereby provides direct information about the DOS and possible many body effects \cite{Wasserman96}.\\ \indent
The magnetization of the MgZnO/ZnO heterostructure at $T=280$~mK for $\alpha=52^{\circ}$ is displayed in Fig.~\ref{fig1}. A smoothly-varying magnetic background signal has been removed from the raw data by subtracting a low-order polynomial fit in $B$. The difference signal $M$ versus $B_{\perp}$ exhibits the dHvA effect as jumps $\Delta M$ in the magnetization occur for $\nu=2$ at $B_{\perp}\approx 3.5$~T (inset) and $\nu=1$ at $B_{\perp}\approx 7$~T. Superimposed at $\nu=1$ is a non-equilibrium (neq) signal $M_{\text{NEC}}$ arising from NECs \cite{Jones1995} showing the characteristic sign change upon changing the magnetic field sweep direction \cite{usher2009,Schwarz2003,Ruhe2009}. Contrary to $\nu=1$, the signal near $\nu=2$ residing at much smaller $B$ does not contain sweep-direction dependent contributions indicating that both the step $\Delta M$ and the spike (marked by an asterix in the inset) reflect equilibrium (eq) properties. An existing (missing) NEC indicates that $\rho_{xx}$ takes a vanishingly small minimum value (remains finite).
\\ \indent
In Fig.~\ref{fig2}~(a) and (b) we depict angular dependencies of the dHvA effect for filling factors $\nu=3$ and $\nu=2$, respectively. In Fig.~\ref{fig2}~(c) we summarize $\Delta M_{\text{e},\nu}=\Delta M_{\nu}/N$ for $\nu=1$ to 3 taken at different $\alpha$. First we discuss $\nu=2$ where spin-polarization of the 2DES is absent. $\Delta M_{\text{e},\nu=2}$ ranges from $1.3~{\rm to}~0.85$~{\mbor} in the angular regime from $\alpha=36.8^{\circ}~{\rm to}~62.0^{\circ}$. Here, \mbor$~=\mu_{\text{B}}(m_{\text{e}}/$\meff) with {\meff}$=0.31(\pm0.01)~m_{\text{e}}$ as determined from a Lifshitz-Kosevich analysis \cite{Shoenberg1984} of $\Delta M(T)$ at $\nu=2$ (not shown). \meff~is consistent with Refs. [\onlinecite{Tsukazaki2007,Tsukazaki2008}]. The corresponding energy gaps $\Delta E_{\nu}=M_{\text{e},\nu}\cdot B_{\perp}$ are shown in Fig.~\ref{fig2}~(d). For an ideal (disorder-free) non-interacting 2DES one would expect the energy gap for $\nu=2$ to amount to $\Delta E_{\nu=2}=\hbar\omega_{\text{c}}-g\mu_{\text{B}}B$. The gap depends on $\alpha$ as denoted by the line in Fig.~\ref{fig2}~(d) because the Landau quantization $\hbar\omega_{\text{c}}=\hbar eB_{\perp}/m^*=\hbar eB\cos\alpha/m^*$ and Zeeman spin splitting energy $|g|\mu_{\text{B}}B$ depend differently on $\alpha$ ($g$ is the band structure Land{\'e} factor). The experimentally determined values of $\Delta E_{\nu=2}$ (triangles) are in good quantitative agreement with the prediction for the ideal 2DES considering $g=1.93$ (line) \footnote{Earlier transport experiments on {\ZnO} heterostructures revealed coincidence phenomena in tilted magnetic fields \cite{Tsukazaki2008,Kozuka2012}. At $\nu=2$, the first coincidence is expected at $\alpha=72.6^\circ$, i.e., beyond the upper limit of the $x$-axis in Fig. \ref{fig2} (d).}. This agreement is surprising as the finite mobility $\mu$ of the 2DES indicates residual disorder in the \ZnO~heterostructure. To further analyze the observed $\Delta M$ we show the dHvA amplitude calculated for the DOS of the real 2DES [Eq. (\ref{Eq1})] in the inset of Fig.~\ref{fig1} as a dotted line. We considered $\Gamma=5\times10^{-2}\sqrt{B_{\perp}}$~meVT$^{-1/2}$. The thereby expected value is indeed smaller than the measured one, indicating that electron-electron interaction enhances the dHvA amplitude in the \ZnO~heterostructure \cite{MacDonald1986}. The dashed line in the inset shows $M(B)$ for the ideal 2DES at $T=0$~K. The expected sawtooth-like shape of the dHvA oscillation is clearly visible. The experimental curve closely follows this behavior indicating the already excellent quality of the 2DES ~\cite{Wiegers1997,Wilde2006}. However, there is an additional spike-like feature (asterix) close to the steep slope. This feature will be discussed later.\\ \indent We now consider the peak-to-peak amplitudes $\Delta M$ of the odd filling factors $\nu=1$ and $\nu=3$ where the Fermi energy resides in the Zeeman energy gap and spin polarisation of the 2DES occurs. We find that $\Delta M_{\text{e},\nu=3}$ varies with $\alpha$ taking a maximum value of $0.68$~{\mbor}. $\Delta M_{\text{e},\nu=1}$ is considerably larger and amounts to a value of up to $\approx2.25~${\mbor}.\\ \indent
From the energy gaps of the odd filling factors $\nu=3$ and $\nu=1$, we extract the Zeeman spin splitting $\Delta E_{\text{Z}}=\left|g^*\right|\mu_{\text{B}}B$. In case of small disorder, spin splitting is known to be enhanced when approaching the quantum limit, i.e., in large fields $B$. This phenomenon is commonly attributed to the exchange interaction $E_{\text{Ex}}$ \cite{MacDonald1986} and expressed in terms of an effective factor $g^*$ \cite{Englert1982} as
\begin{equation}
\Delta E_{\text{Z}}=\left|g\right|\mu_{\text{B}}B+E_{\text{Ex}}=\left|g^*\right|\mu_{\text{B}}B.
\end{equation}
From the data shown in Fig.~\ref{fig2}~(d) we extract a maximum $g^*=1.1$ for $\nu=3$. This value is reduced compared with the band structure $g$ factor of ZnO. We attribute the reduction of $g^*$ to the level broadening and in particular level overlap in the low field regime. For $\nu=1$, we find a maximum $g^*$ of $5.1$ at $\alpha=36.8~$deg. This value is significantly enhanced over $g=1.93$ \footnote{In Ref.~\onlinecite{Kozuka2012}, $g^*=7.28$ was reported when addressing excited states in a transport experiment.}. Our dHvA experiment thus substantiates that electron-electron interaction plays an important role in the magnetic field dependent ground state properties of the {\ZnO} heterostructure at both an even and odd integer $\nu$.
\begin{figure}[htbp]
\center
\includegraphics[width=7cm]{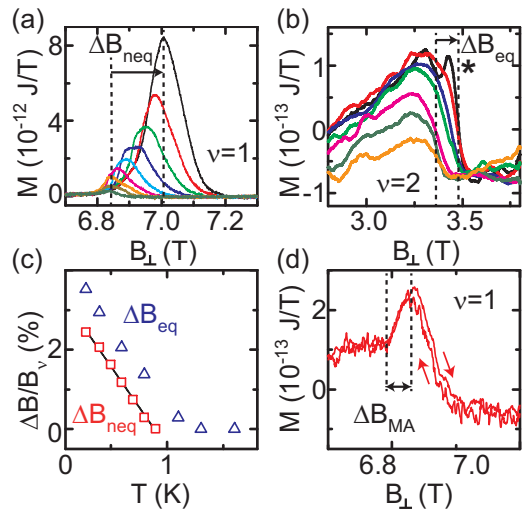}
\caption{(Color online) Temperature dependence of $M(B)$ near (a) $\nu=1$ at $\alpha=36.8^{\circ}$ and (b) $\nu=2$ at $\alpha=62.0^{\circ}$ measured for down-sweeping $B$. From top to bottom the temperature varies from $T=0.28~{\rm to}~1.6$~K. For $T\leq0.9$~K NECs are present at $\nu=1$ in (a). Going from higher to lower temperatures, the field position of the maximum non-equilibrium signal in (a) moves to larger $B$ by $\Delta B_{\rm neq}$. A shift $\Delta B_{\rm eq}$ is seen also for the equilibrium signal in (b). (c) $\Delta B_{\rm neq}/B_{\nu=1}$ (squares) and $\Delta B_{\rm eq}/B_{\nu=2}$ (triangles) as a function of $T$. At low $T$, $\Delta B_{\rm neq}$ increases linearly with decreasing $T$ as indicated by a linear fit (black line). (d) Spike-like overshoot of $M(B)$ at $\nu=1$ and $T=1.6$~K.}
\label{fig3}
\end{figure}\\ \indent
Figure~\ref{fig3}~(a) and (b) shows the temperature dependence of $M(B)$ near $\nu=1$ and $\nu=2$, respectively, in the temperature range $0.28$ to $1.6$~K. For $T\leq0.9$~K, there is a large signal arising from NECs superimposed on the dHvA signal at $\nu=1$. The peak value of the NEC decreases with increasing temperature as already reported for GaAs/AlGaAs heterostructures \cite{Jones1995}. Here, we focus on the field position of the NEC maxima. At this position, the Fermi energy resides in localized states and the QHE edge channels are most strongly decoupled from the bulk, resulting in an effective suppression of backscattering. NECs are a very sensitive gauge for this effect, since the breakdown-limited NEC signal increases strongly for increasing suppression of backscattering. This is in contrast to magnetotransport experiments, where $\rho_{xx}$ and $\rho_{xy}$ exhibit extended zeros and plateaus, respectively. Indeed, from the time-dependent decay of NECs, resistivities $<10^{-14}$~$\Omega/\Box$ have been inferred \cite{usher2009}. Interestingly we find that the temperature-dependent maximum value shifts towards smaller magnetic field values with increasing $T$. The shift of the non-equilibrium magnetization is highlighted for $\nu=1$ in Fig.~\ref{fig3}~(c) where we plot the relative separation $\Delta B_{\rm neq}/B_{\nu=1}$ between peaks at low $T$ and the peak at $T=0.9~$K (squares). The peak position varies almost linearly with $T$ (line). Following Ref.~\onlinecite{Tsukazaki2007} we exclude that the observed shifts arise from temperature dependent variations in $n_{\text{s}}$ as they set in only for $T>1~$K. For $\nu=2$ [Fig.~\ref{fig3}~(b)], we do not observe NECs in the temperature regime studied. The spike-like overshoot (asterix) is an equilibrium feature. However, we find a shift of the equilibrium dHvA oscillation to smaller $B$ for increasing $T$. The relative shift $\Delta B_{\rm eq}/B_{\nu=2}$ is displayed in Fig.\ \ref{fig3}~(c). Below $\sim 1$~K, it also increases linearly as $T$ decreases.
\\ \indent We now discuss the temperature dependent shifts of non-equilibrium and equilibrium features. As mentioned above, the field position of the maximum NEC is directly connected to the minimum of $\rho_{{xx}}$ \cite{Ruhe2009}, where backscattering is maximally suppressed. Our data suggest that the minima of $\rho_{\text{xx}}$ shift towards higher magnetic fields with decreasing $T$, i.e. towards the high energy side of the energy spectrum. Also the field position of the dHvA oscillation at $\nu=2$ shifts by $\Delta B_{\rm eq}$ as a function of $T$. The relative shift of both, $\Delta B_{\rm neq}$ and $\Delta B_{\rm eq}$, is consistent in temperature and we thus suppose that they have the same microscopic origin. For an equilibrium property such as the dHvA effect, no temperature dependent shifts have been reported before. Concerning non-equilibrium properties, displacements and asymmetries of QHE plateaus and corresponding $\rho_{{xx}}$ minima have been observed in transport measurements on GaAs-based heterostructures \cite{Furneaux1986,Haug1987,Raymond2009}. They were attributed to scattering centers being present in or in close vicinity to the 2DES. Repulsive scattering centers altered the DOS in such a way that it became asymmetric exhibiting an impurity tail at its high energy side \cite{Bonifacie2006}, while attractive scatterers led to the opposite asymmetry \cite{Raymond2009}. As a consequence, localized states of $\nu=2$ and extended states of $\nu=1$ in the DOS could overlap resulting in a shift of $\rho_{\text{xx}}$ minima for repulsive scatterers. We find corresponding displacements in transport measurements on a reference sample \cite{supmaterial}. From the direction of the observed shift in field position we infer that the interaction of charge carriers and scatterers in the \ZnO~heterostructure is repulsive.\\ \indent
We further observed spike-like overshoots of the equilibrium magnetization as depicted in the inset of Fig.~\ref{fig1} and Fig. \ref{fig3} (b) for $\nu=2$ at 0.28~K. Regarding $\nu=1$, we also observe an overshoot which is depicted in Fig.~\ref{fig3}~(d) at $T=1.6$~K, where NECs are absent. Here, the small remaining hysteresis is attributed to the self-induction of the superconducting magnet and the finite averaging time for data acquisition. The overshoots do not depend on the field sweep direction. Similar effects were reported for a bi-layered 2DES in GaAs/AlGaAs \cite{Bominaar2006b}. The authors speculated that correlations effects or a peculiar form of the DOS were responsible for the overshoot. In light of the fact of an asymmetric DOS we attribute this feature to magnetic thaw down of electrons \cite{Bisotto2012}. In the presence of repulsive scatterers in close vicinity to the 2DES, localized magnetoacceptor states can exist at energies above the extended Landau states \cite{Bisotto2012}. When electrons are transferred from the higher-lying localized to the lower-lying extended states, the free energy $F$ of the electron system decreases. Correspondingly, there is an increase in $M=-\frac{\partial F}{\partial B}$ enhancing the equilibrium dHvA amplitude with a spike-like overshoot in the magnetic thaw down process near an integer $\nu$. We apply the formalism of Ref.~\onlinecite{Wiegers1997} and estimate the average density of magnetoacceptors by $n_{\text{MA}}=n_{\text{s}}\Delta B_{\text{MA}}/B_{\text{av}}$, where $\Delta B_{\text{MA}}$ is defined in Fig.~\ref{fig3}~(d). $B_{\text{av}}$ is the mean field value. We get $n_{\text{MA}}\approx10^{9}~$cm$^{-2}$. Magnetization measurements offer thus a tool to determine the nature and number of scatterers and thereby getting detailed insight into the microscopic sample structure. This might help to further optimize MgZnO/ZnO heterostructures. One may now speculate about the origin of the repulsive scatterers in the vicinity of the 2DES in the {\ZnO} heterostructure. A possible explanation for repulsive scattering centers is the presence of Zn vacancies at the interface, which has also been reported in Ref.~\cite{uedono2003}.\\ \indent
In summary, we studied correlation and disorder effects in the {\ZnO} heterostructure by means of the dHvA effect and NECs. The energy gaps at $\nu=2$ and $\nu=1$ were enhanced over the expected value in the single particle picture which we ascribe to exchange interaction in the dilute 2DES. Temperature dependent shifts of the filling factors and of the maxima of the NECs were found indicating the presence of repulsive scattering centers in the direct vicinity of the 2DES. The shape of the dHvA oscillations was consistent with magnetic thaw down from magnetoacceptor states enhancing the quantum oscillatory magnetization even further beyond the electron-electron interaction.\\ \indent We acknowledge financial support from the Deutsche Forschungsgemeinschaft via TRR80. This work was partly supported by Grant-in-Aids for Scientific Research (S) No. 24226002 from MEXT, Japan as well as by Murata Science Foundation (Y.K.).

\clearpage
\onecolumngrid

\section{\Large{Supplemental Material}}
%Enhanced quantum oscillatory magnetization in an oxide heterostructure due to electron-electron interaction and disorder by repulsive scatterers}% Force line breaks with \\
\hspace{2cm}
\section{Sample details}

\begin{figure}[htbp]
\center
\includegraphics[width=8cm]{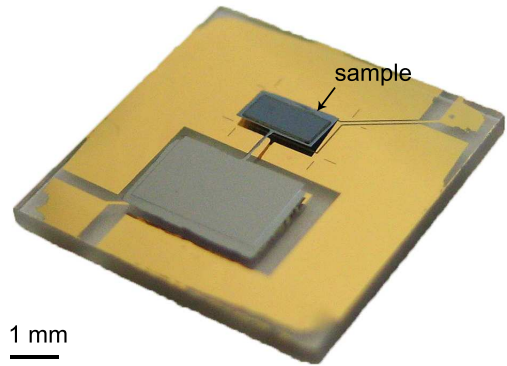}
\caption[width=6cm]{Photo of a thinned {\ZnO} sample on micromechanical cantilever sensor prepared from an undoped AlGaAs/GaAs heterostructure. The active {\ZnO} 2DES area is 1.8$\times$0.9 mm$^2$. }
\label{fig4}
\end{figure}

The heterostructure was grown using molecular beam epitaxy with high-purity 7 N Zn and 6 N Mg metals as well as distilled pure ozone as the oxygen source. The heterostructure was formed by depositing a 690~nm-thick ZnO layer followed by a 380~nm-thick MgZnO layer on a Zn-polar ZnO substrate. A 2DES is spontaneously formed at the interface without remote doping of donors due to the polarization mismatch between the ZnO and MgZnO layers. To allow for cantilever magnetometry experiments, the sample was cut by a wiresaw and polished from the backside to reduce the thickness of the substrate to about $30$~$\mu$m. Figure ~\ref{fig4} shows a photograph of a thinned MgZnO/ZnO sample attached to a GaAs-based micromechanical cantilever magnetometer. The cantilever is glued to an Au-coated sapphire substrate containing the contact pads and guard plane for the capacitive readout \cite{Wilde2008}.

\section{Transport characterization}

\begin{figure}[htbp]
\center
\includegraphics[width=12cm]{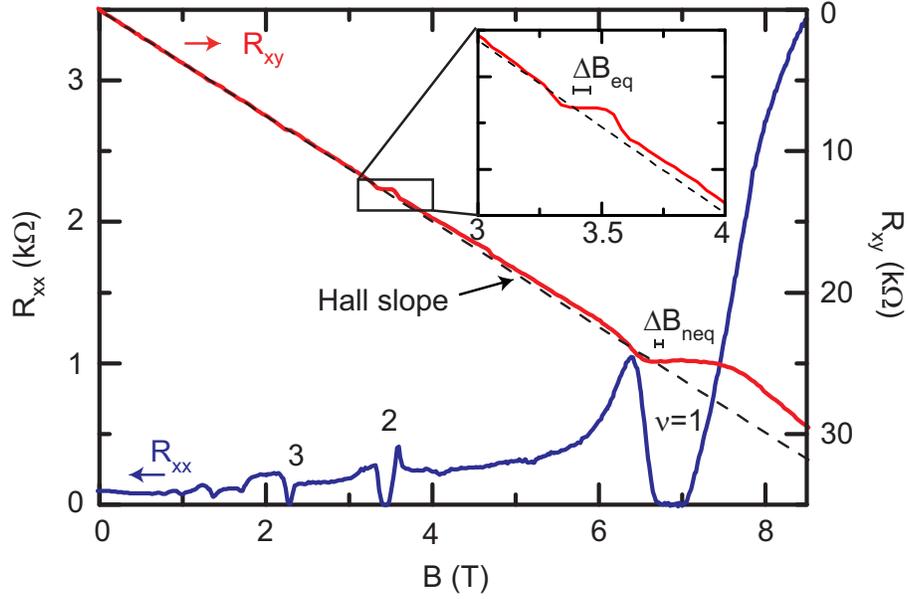}
\caption[width=9cm]{Transport characterization of the {\ZnO} reference sample A at $T=0.5$~K yields $n_{\text{s}}=1.7\times10^{11}$~cm$^{-2}$ and $\mu=3.8\times10^5$~cm$^2$/Vs. The crossing of the Hall slope with the Hall plateaus at $\nu=1$ and $\nu=2$ does not occur in the center of the respective plateau. Instead, the plateaus as well as the $R_{\text{xx}}$ minima are shifted to larger fields. The shifts $\Delta B_{\rm neq}$ and $\Delta B_{\rm eq}$ observed in the magnetometry experiment are indicated.}
\label{fig5}
\end{figure}

Transport experiments have been conducted on a reference sample cut from the same heterostructure at $T=0.5$~K using the van-der-Pauw method. We refer to this sample as reference sample A in the following. The results for $R_{\text{xx}}$ (blue line) and $R_{\text{xy}}$ (red line) are shown in Fig.~\ref{fig5}. Analyzing the Shubnikov-de Haas oscillations and the Hall effect yields $n_{\text{s}}=1.7\times10^{11}$~cm$^{-2}$ and $\mu=3.8\times10^5$~cm$^2$/Vs. The classical Hall slope extrapolated from the low field data is shown by the dashed line. At $\nu=2$ and $\nu=1$, the  Hall plateaus in $R_{\text{xy}}$  do not occur symmetrically with respect to the classical Hall slope. Instead, the Hall plateaus as well as the $R_{\text{xx}}$ minima are shifted to larger fields. Following Refs.~\onlinecite{Furneaux1986,Haug1987,Raymond2009}, this behavior is attributed to the presence of repulsive scatterers being in close vicinity to the 2DES.

\section{Magnetization of a reference 2DES}

\begin{figure}[htbp]
\center
\includegraphics[width=11cm]{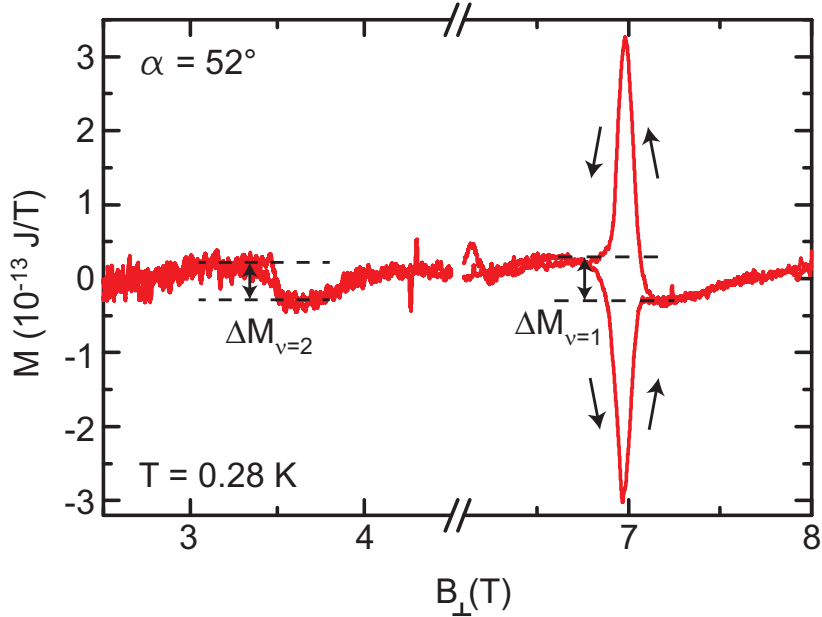}
\caption[width=9cm]{Magnetization as a function of $B_{\perp}$ of reference sample B at $T=0.28$~K and $\alpha=52^{\circ}$. At $\nu=2$ and $\nu=1$, the data exhibits the dHvA effect. The amplitudes per electrons are $\Delta M_{\text{e},\nu=2}=0.6$~$\mu_{\text{B}}^*$ and $\Delta M_{\text{e},\nu=1}=0.8$~$\mu_{\text{B}}^*$, respectively. NECs are resolved near $\nu=1$. The amplitude of $M_{\text{NEC}}$ amounts to $\sim 3\times10^{-13}$~J/T.}
\label{fig6}
\end{figure}
In order to establish whether the observed magnetization features occur systematically, additional magnetization experiments were conducted, following the measurement procedure described in the main text. Here, a second 2DES sample cut from the same heterostructure was used, which we will refer to as reference sample B in the following. The result of the experiment at $T=0.28$~K and $\alpha=52^{\circ}$ is depicted in Fig.~\ref{fig6}. The jumps in the magnetization at $\nu=2$ and $\nu=1$ can be ascribed to the dHvA effect. A signal stemming from the NECs, which changes sign upon changing sweep direction, is present at $\nu=1$. The behavior as a function of temperature of both, dHvA effect and NECs, is consistent with the one described in the main text (not shown). The NEC maxima at $\nu=1$ and the position of the dHvA oscillation at $\nu=2$ shift systematically towards lower fields with increasing temperature. In comparison to the results shown in the main text, the observed signal amplitudes of the dHvA effect and the NECs are smaller here. Regarding the dHvA effect, we observe $\Delta M_{\text{e},\nu=2}=0.6$~$\mu_{\text{B}}^*$ and $\Delta M_{\text{e},\nu=1}=0.8$~$\mu_{\text{B}}^*$. These values are by a factor of 2 to 3 smaller than the ones reported in the main text. The dHvA effect at $\nu=3$ was not resolved. The signal arising from the non-equilibrium magnetization $M_{\text{NEC}}$ is reduced by a factor of 10.\\
The reduced signal amplitudes in the reference sample B might be ascribed to variations in the 2DES homogeneity and mobility within the {\ZnO} heterostructure such that the 2DES properties vary more strongly in reference sample B. It is well established that the dHvA effect is very sensitive to disorder and sample inhomogeneity \cite{Eisenstein1985,Wilde2008}. Their mutual effect is to wash out the dHvA oscillations. However, we clearly observe consistent temperature dependent shifts of equilibrium- as well as non-equilibrium magnetization signals on the $B$-field axis also in the reference sample B, \emph{despite} the apparent differences in the sample parameters. This indicates that the observed phenomena are quite robust.

\end{document}